\documentclass[12pt]{article}
\usepackage{graphicx} % Required for inserting images
\usepackage{amsmath}
\usepackage{natbib}
\usepackage{helvet}
\usepackage{fancyhdr}
\usepackage{mathbbol}
\usepackage[margin=1in]{geometry}  % This sets 1 inch margins on all sides

\title{Bayesian algorithmic perfumery: \\ A Hierarchical Relevance Vector Machine for the Estimation of Personalized Fragrance Preferences based on Three Sensory Layers and Jungian Personality Archetypes}
\author{Rolando Gonzales Martinez
\medskip \\
        \textsf{University of Groningen} \\
\texttt{r.m.gonzales.martinez@rug.nl}}
\date{November 2024}

\begin{document}

\maketitle
\thispagestyle{empty} 

\begin{abstract}
\noindent This study explores a Bayesian algorithmic approach to personalized fragrance recommendation by integrating hierarchical Relevance Vector Machines (RVM) and Jungian personality archetypes. The paper proposes a structured model that links individual scent preferences for top, middle, and base notes to personality traits derived from Jungian archetypes, such as the Hero, Caregiver, and Explorer, among others. The algorithm utilizes Bayesian updating to dynamically refine predictions as users interact with each fragrance note. This iterative process allows for the personalization of fragrance experiences based on prior data and personality assessments, leading to adaptive and interpretable recommendations. By combining psychological theory with Bayesian machine learning, this approach addresses the complexity of modeling individual preferences while capturing user-specific and population-level trends. The study highlights the potential of hierarchical Bayesian frameworks in creating customized olfactory experiences, informed by psychological and demographic factors, contributing to advancements in personalized product design and machine learning applications in sensory-based industries.
\end{abstract}

\newpage
\section{Introduction}
The development of personalized fragrance recommendations requires an understanding of individual personality traits, preferences, and sensory responses. Jungian psychology, with its focus on personality archetypes and cognitive functions, provides a useful framework for interpreting individual differences in preferences, including olfactory preferences, which are central in perfumery. Jung’s concept of archetypes describes fundamental character types that represent universal, recurring patterns in human personality and behavior \citep{jung1959archetypes}. These archetypes, such as the Hero, Caregiver, Explorer, and Lover, can influence sensory inclinations, suggesting that personality might drive fragrance preferences as well \citep{feist2021theories}.

The science of perfumery has historically focused on the art of blending aromatic compounds to create layered scents that evolve over time. Typically, fragrances are structured in three sequential notes—top, middle, and base—which unfold gradually to provide a full olfactory experience \citep{ackerman2006smell}. Each note layer has distinct characteristics, and users’ preferences for specific notes may reflect their personality traits, as discussed in Jungian psychological models.

Modern machine learning approaches, particularly Bayesian methods, allow for personalization in recommendations by dynamically updating individual preferences based on new observations. Bayesian methods are well-suited for handling uncertainty and can adaptively update predictions based on evidence \citep{bishop2006pattern}. A Bayesian approach allows for modeling complex, hierarchical relationships where both individual and group-level trends influence outcomes. Hierarchical Bayesian models (HBMs) provide a powerful way to structure these relationships, capturing both individual and population-level variation \citep{gelman2013bayesian}.

In machine learning, Bayesian algorithms like the Relevance Vector Machine (RVM) extend Bayesian methods to classification and regression tasks by using sparse probabilistic models. The RVM, which includes sparsity-inducing priors, can capture relevant patterns with fewer data points, making it ideal for applications where interpretability and personalized predictions are key \citep{tipping2001sparse}. For personalized fragrance preferences, a hierarchical Bayesian RVM can integrate Jungian archetypes as personality priors, updating predictions as users interact with different fragrance notes. This combination of Jungian psychology, perfumery, and Bayesian machine learning offers a structured, adaptive approach to understanding and predicting olfactory preferences based on individual personality traits.

Perfume is typically crafted in three layers of \textit{notes} that create a full sensory experience as the fragrance evolves. These layers are top notes, middle (or heart) notes, and base notes, each with distinct characteristics and functions. Together, these three notes create the full fragrance profile that changes subtly over time.

Top notes are the initial scents you smell right after applying the perfume. They are often light, fresh, and evaporate quickly, lasting around 15–30 minutes. Top notes make the first impression, usually including citrus, herbs, or light floral elements like lemon, lavender, or mint.

Middle (heart) notes develop as the top notes fade, forming the core of the fragrance. Middle notes last longer than top notes, often from 1–5 hours, and create the perfume's main body. Heart notes are usually warmer and fuller, like rose, jasmine, or spices, balancing the top notes and harmonizing with the deeper base notes.

Base Note are the foundation of the perfume, lingering for hours or even all day. Base notes are usually rich and long-lasting, grounding the fragrance with depth and warmth. They often include ingredients like sandalwood, musk, amber, or vanilla, leaving a lasting impression. 

\section{A Bayesian approach to perfumery}

A Bayesian updating rule that links the three perfume notes---top, middle, and base---can be created by considering each note as a sensory hypothesis that contributes to the overall perception of the fragrance over time. A framework that incorporates Bayesian updating to assess the evolving probability that the fragrance will be perceived as ``pleasant'' (or another positive sensory quality) as it transitions through its notes, can be defined by following 3 steps.

% Step 1: Define Sensory Hypotheses for Each Note
Let:
\begin{itemize}
    \item \( H_T \): Hypothesis that the top note is perceived as pleasant.
    \item \( H_M \): Hypothesis that the middle note is perceived as pleasant.
    \item \( H_B \): Hypothesis that the base note is perceived as pleasant.
\end{itemize}

Our goal is to find the probability that the overall fragrance experience (\( H_F \)) will be pleasant, given the perceptions of each note.

% Step 2: Establish Prior Probabilities
Before any note is perceived, assign initial prior probabilities based on expectations of each note's pleasantness, derived from historical data, customer feedback, or composition expertise:
\begin{itemize}
    \item \( P(H_T) \): Prior probability that the top note is pleasant.
    \item \( P(H_M) \): Prior probability that the middle note is pleasant.
    \item \( P(H_B) \): Prior probability that the base note is pleasant.
\end{itemize}

% Step 3: Update Probabilities Sequentially Using Bayesian Updating
As each note is perceived, update the probability of the overall fragrance experience being pleasant.

1. \textbf{Top Note}:\\
Upon perceiving the top note, we update our belief about the pleasantness of the fragrance as follows:
\[
   P(H_F | H_T) = \frac{P(H_T | H_F) \cdot P(H_F)}{P(H_T)}
\]
where:
\begin{itemize}
    \item \( P(H_F | H_T) \): Posterior probability of overall pleasantness after the top note is experienced.
    \item \( P(H_T | H_F) \): Likelihood of perceiving a pleasant top note given a pleasant fragrance.
\end{itemize}

2. \textbf{Middle Note}:\\
After experiencing the middle note, update again:
\[
   P(H_F | H_T, H_M) = \frac{P(H_M | H_F, H_T) \cdot P(H_F | H_T)}{P(H_M)}
\]
Here, the probability is adjusted given the experience of both top and middle notes.

3. \textbf{Base Note}:\\
Finally, update with the base note to finalize the probability of the fragrance being perceived as pleasant:
\[
   P(H_F | H_T, H_M, H_B) = \frac{P(H_B | H_F, H_T, H_M) \cdot P(H_F | H_T, H_M)}{P(H_B)}
\]
This cumulative probability now reflects the perception after experiencing all three notes.

% Interpretation
The final probability, \( P(H_F | H_T, H_M, H_B) \), reflects the likelihood that the fragrance is perceived as pleasant after sequentially experiencing each note. Each note contributes to adjusting this probability, allowing for a dynamic, evolving assessment of the fragrance’s overall appeal.

\section{Bayesian Machine Learning Algorithm for Personalized Fragrance Preferences}

We aim to design a Bayesian machine learning algorithm that integrates each person's prior preference for fragrance notes (top, middle, base) with data-driven evidence based on their personality traits, as determined by Jungian personality tests. This approach allows us to combine an individual’s prior "taste" with their personality profile, producing a customized fragrance recommendation model.

% Step 1: Define the Problem and Variables

Let:
\begin{itemize}
    \item \( \theta_i \): Each person \( i \)'s true preference for a specific fragrance note.
    \item \( D_i \): Observed data, including responses from a Jungian personality test and any previous fragrance ratings by person \( i \).
    \item \( P(\theta_i) \): Prior probability distribution representing \( i \)'s initial preference for fragrance notes.
    \item \( P(D_i | \theta_i) \): Likelihood function representing how likely we are to observe \( D_i \) given a particular preference \( \theta_i \).
    \item \( P(\theta_i | D_i) \): Posterior probability distribution that combines \( i \)'s prior preference and personality test data, providing an updated fragrance preference profile.
\end{itemize}

% Step 2: Establish Prior Preferences Based on Fragrance Notes

Each person’s prior preference \( P(\theta_i) \) for fragrance notes is initially estimated using demographic data, historical preferences, or a short initial survey on scent preferences. This prior distribution reflects the individual’s initial "taste" before introducing personality-based evidence.

% Step 3: Define the Likelihood Function Based on Personality Data

We introduce Jungian personality factors, such as introversion/extroversion and intuition/sensation, as evidence that correlates with fragrance preferences. For example:
\begin{itemize}
    \item Extroverts may prefer more intense, outgoing fragrance notes.
    \item Introverts may lean toward subtle or natural scents.
\end{itemize}
The likelihood function \( P(D_i | \theta_i) \) represents the probability of observing \( i \)'s personality data given specific fragrance preferences.

% Step 4: Bayesian Updating Rule to Calculate Posterior Preference

Using Bayes' theorem, we update each person’s preference as follows:
\[
   P(\theta_i | D_i) = \frac{P(D_i | \theta_i) \cdot P(\theta_i)}{P(D_i)}
\]
where:
\begin{itemize}
    \item \( P(\theta_i | D_i) \): Posterior distribution, the updated fragrance preference for person \( i \) after considering personality test data.
    \item \( P(D_i | \theta_i) \): Likelihood of observing \( i \)'s personality data given their preference.
    \item \( P(\theta_i) \): Prior fragrance preference.
    \item \( P(D_i) \): Marginal likelihood, serving as a normalizing constant.
\end{itemize}

% Step 5: Iterative Update and Predictive Model

Once the posterior \( P(\theta_i | D_i) \) is calculated, it serves as an updated preference profile. This profile can be iteratively updated with new data points, such as additional personality insights or fragrance feedback, using the same Bayesian updating process.

The algorithm can then predict which fragrances are likely to be well-received by \( i \), given their personality-adjusted preferences, helping create a personalized fragrance experience.

% Bayesian Machine Learning Algorithm with Update Rule for Personalized Fragrance Preferences Based on 3 Notes

This Bayesian machine learning algorithm integrates each person’s prior taste for fragrance notes (top, middle, base) with evidence from their personality, calculated from Jungian personality tests. The model evolves dynamically with Bayesian updating at each note level, refining predictions of a person's overall fragrance preference as each note is experienced.

% Step 1: Define the Problem and Variables

Let:
\begin{itemize}
    \item \( \theta_i \): Each person \( i \)'s true preference for a specific fragrance note.
    \item \( D_i \): Observed data, including responses from a Jungian personality test and any previous fragrance ratings by person \( i \).
    \item \( P(\theta_i) \): Prior probability distribution representing \( i \)'s initial preference for fragrance notes.
    \item \( P(D_i | \theta_i) \): Likelihood function representing how likely we are to observe \( D_i \) given a particular preference \( \theta_i \).
    \item \( P(\theta_i | D_i) \): Posterior probability distribution that combines \( i \)'s prior preference and personality test data, providing an updated fragrance preference profile.
\end{itemize}

% Step 2: Establish Prior Preferences Based on Fragrance Notes

Each person’s prior preference \( P(\theta_i) \) for fragrance notes is initially estimated using demographic data, historical preferences, or a short initial survey on scent preferences. This prior distribution reflects the individual’s initial "taste" before introducing personality-based evidence.

% Step 3: Define the Likelihood Function Based on Personality Data

We introduce Jungian personality factors, such as introversion/extroversion and intuition/sensation, as evidence that correlates with fragrance preferences. For example:
\begin{itemize}
    \item Extroverts may prefer more intense, outgoing fragrance notes.
    \item Introverts may lean toward subtle or natural scents.
\end{itemize}
The likelihood function \( P(D_i | \theta_i) \) represents the probability of observing \( i \)'s personality data given specific fragrance preferences.

% Step 4: Bayesian Updating Rule with 3 Notes

We sequentially update the overall fragrance preference as each note (top, middle, base) is experienced. This is done using Bayes' theorem at each stage:

1. Top Note:\\
   After experiencing the top note, update the probability of overall fragrance preference as follows:
   \[
      P(\theta_i | H_T) = \frac{P(H_T | \theta_i) \cdot P(\theta_i)}{P(H_T)}
   \]
   where:
   \begin{itemize}
       \item \( P(\theta_i | H_T) \): Posterior probability of preference after experiencing the top note.
       \item \( P(H_T | \theta_i) \): Likelihood of perceiving the top note as pleasant, given the person’s preference.
   \end{itemize}

2. Middle Note:\\
   Upon perceiving the middle note, update the probability further:
   \[
      P(\theta_i | H_T, H_M) = \frac{P(H_M | \theta_i, H_T) \cdot P(\theta_i | H_T)}{P(H_M)}
   \]
   Here, the probability reflects both top and middle note experiences.

3. Base Note:\\
   After experiencing the base note, update the final probability of preference for the fragrance:
   \[
      P(\theta_i | H_T, H_M, H_B) = \frac{P(H_B | \theta_i, H_T, H_M) \cdot P(\theta_i | H_T, H_M)}{P(H_B)}
   \]
   This cumulative probability, \( P(\theta_i | H_T, H_M, H_B) \), reflects the person’s updated preference profile based on all three notes.

% Step 5: Iterative Update and Predictive Model

The final posterior \( P(\theta_i | H_T, H_M, H_B) \) represents a dynamically personalized fragrance profile for person \( i \) after sequentially experiencing each note. This model can be continuously refined with new observations, such as additional personality data or fragrance preferences, making it adaptable to individual changes over time.

% Ideal Bayesian Machine Learning Algorithm for Personalized Fragrance Preferences Based on 3 Notes: Hierarchical Bayesian Model (HBM)

A Hierarchical Bayesian Model (HBM) is ideal for modeling personalized fragrance preferences based on sequential notes. HBMs allow us to model individual preferences while leveraging population-level trends, making it possible to personalize experiences and utilize shared patterns in fragrance preferences. This is particularly useful in combining unique user preferences with sequential note updates.

% Step 1: Hierarchical Structure for Individual and Population-Level Insights

Define:
\begin{itemize}
    \item \textbf{Individual-Level Preferences}: Each user \( i \) has a unique preference for top, middle, and base notes, represented by parameters \( \theta_i^{(T)}, \theta_i^{(M)}, \theta_i^{(B)} \). These parameters capture the individual user’s subjective fragrance experience.
    \item \textbf{Population-Level Trends}: At the population level, there exist common trends across users, represented by hyperparameters \( \mu \) (mean preference) and \( \sigma \) (variance) for each note. These hyperparameters allow the model to capture broader relationships between personality traits and fragrance preferences.
\end{itemize}

The individual parameters are drawn from population-level distributions, so:
\[
\theta_i^{(T)}, \theta_i^{(M)}, \theta_i^{(B)} \sim \mathcal{N}(\mu, \sigma^2)
\]
where \( \mu \) and \( \sigma \) are inferred from data and describe the typical population-level preferences for each fragrance note.

% Step 2: Sequential Bayesian Updating with Notes

We use Bayesian updating to refine each user’s preference as they experience the top, middle, and base notes in sequence. This allows for a dynamic model where the posterior from each note experience becomes the prior for the next.

% Step 3: Define the Priors and Likelihood Function

\textbf{Priors}: We initialize priors for each user’s preferences based on demographic or survey data, representing the likelihood of each note’s pleasantness.

Let:
\begin{itemize}
    \item \( P(\theta_i^{(T)} | \mu, \sigma^2) \): Prior distribution for the top note preference for user \( i \), based on population mean \( \mu \) and variance \( \sigma^2 \).
    \item \( P(\theta_i^{(M)} | \theta_i^{(T)}, \mu, \sigma^2) \): Prior for the middle note, updated based on the top note preference.
    \item \( P(\theta_i^{(B)} | \theta_i^{(T)}, \theta_i^{(M)}, \mu, \sigma^2) \): Prior for the base note, updated based on both top and middle note experiences.
\end{itemize}

\textbf{Likelihood Function}: The likelihood of perceiving each note as pleasant is influenced by personality covariates (e.g., Jungian traits). For example:
\[
P(D_i | \theta_i^{(T)}, \theta_i^{(M)}, \theta_i^{(B)}) = f(\text{personality covariates})
\]
where \( D_i \) represents the observed data for user \( i \), including their personality profile.

% Step 4: Bayesian Updating Rule with 3 Notes

For each note, we apply Bayesian updating, incorporating the pleasantness experience of the note into the user’s preference profile.

1. Top Note:
   \[
   P(\theta_i^{(T)} | D_i^{(T)}) = \frac{P(D_i^{(T)} | \theta_i^{(T)}) \cdot P(\theta_i^{(T)} | \mu, \sigma^2)}{P(D_i^{(T)})}
   \]
   where:
   \begin{itemize}
       \item \( P(\theta_i^{(T)} | D_i^{(T)}) \): Posterior distribution for the top note preference after observing user \( i \)'s experience with the top note.
       \item \( P(D_i^{(T)} | \theta_i^{(T)}) \): Likelihood of perceiving the top note as pleasant.
       \item \( P(\theta_i^{(T)} | \mu, \sigma^2) \): Prior probability for the top note preference.
   \end{itemize}

2. Middle Note:
   \[
   P(\theta_i^{(M)} | D_i^{(T)}, D_i^{(M)}) = \frac{P(D_i^{(M)} | \theta_i^{(M)}, \theta_i^{(T)}) \cdot P(\theta_i^{(M)} | \theta_i^{(T)}, \mu, \sigma^2)}{P(D_i^{(M)})}
   \]
   Here, \( P(\theta_i^{(M)} | D_i^{(T)}, D_i^{(M)}) \) represents the posterior preference for the middle note, accounting for the top note experience.

3. Base Note:
   \[
   P(\theta_i^{(B)} | D_i^{(T)}, D_i^{(M)}, D_i^{(B)}) = \frac{P(D_i^{(B)} | \theta_i^{(B)}, \theta_i^{(T)}, \theta_i^{(M)}) \cdot P(\theta_i^{(B)} | \theta_i^{(T)}, \theta_i^{(M)}, \mu, \sigma^2)}{P(D_i^{(B)})}
   \]
   where \( P(\theta_i^{(B)} | D_i^{(T)}, D_i^{(M)}, D_i^{(B)}) \) is the posterior preference for the base note after accounting for all prior notes.

% Step 5: Final Posterior Preference

After sequentially updating preferences for each note, we obtain the final posterior \( P(\theta_i^{(B)} | D_i^{(T)}, D_i^{(M)}, D_i^{(B)}) \), which reflects a personalized fragrance preference profile for user \( i \), considering each note’s experience and population-level parameters.

\[
P(\theta_i | D_i) = P(\theta_i^{(B)} | D_i^{(T)}, D_i^{(M)}, D_i^{(B)})
\]

This final posterior represents a tailored fragrance prediction, allowing for an adaptive model that evolves based on both individual and population-level insights. The model can be further refined by incorporating additional personality and preference data for continued personalization.

% Hierarchical Bayesian Model for Personalized Fragrance Preferences Using Relevance Vector Machine (RVM) Mathematics

We can use the mathematics of a Relevance Vector Machine (RVM) to build a Hierarchical Bayesian Model (HBM) for personalized fragrance preferences. An RVM provides a probabilistic framework with sparsity-inducing priors that can capture both individual (user-specific) and population-level fragrance preferences. This approach is effective for sequential updates with each note (top, middle, and base) while considering individual personality traits and prior preferences.

% Step 1: Define the Model Components

Let:
\begin{itemize}
    \item \( \mathbf{y}_i \): The target fragrance preference score for user \( i \), observed after experiencing all three notes.
    \item \( \mathbf{x}_i = [x_i^{(T)}, x_i^{(M)}, x_i^{(B)}] \): Feature vector for user \( i \), incorporating Jungian personality traits, demographics, and prior fragrance preferences for top, middle, and base notes.
    \item \( \boldsymbol{\phi}(\mathbf{x}_i) \): Basis function to map inputs to a high-dimensional feature space for capturing non-linear preferences.
    \item \( \mathbf{w} \): Weight vector for the basis functions, representing the strength of each feature.
\end{itemize}

The goal is to estimate the probability of a target preference \( y_i \) (pleasant or unpleasant) for each user given their features \( \mathbf{x}_i \).

% Step 2: Likelihood Function

In an RVM, the likelihood of observing the target preference score \( \mathbf{y}_i \) for user \( i \) given the weight vector \( \mathbf{w} \) and basis functions is modeled as:
\[
P(\mathbf{y}_i | \mathbf{w}, \boldsymbol{\phi}(\mathbf{x}_i), \sigma^2) = \mathcal{N}(\mathbf{y}_i | \mathbf{w}^T \boldsymbol{\phi}(\mathbf{x}_i), \sigma^2)
\]
where:
\begin{itemize}
    \item \( \mathbf{w}^T \boldsymbol{\phi}(\mathbf{x}_i) \) is the prediction of the fragrance preference for user \( i \).
    \item \( \sigma^2 \): Variance, representing noise in the target preference score.
\end{itemize}

% Step 3: Priors for Individual-Level Weights (Sparsity-Inducing Priors)

To induce sparsity, assign an individual Gaussian prior over each weight \( w_j \) for feature \( j \):
\[
P(w_j | \alpha_j) = \mathcal{N}(w_j | 0, \alpha_j^{-1})
\]
where \( \alpha_j \) is a hyperparameter that controls the precision (inverse variance) of each weight \( w_j \). Large values of \( \alpha_j \) drive \( w_j \) toward zero, allowing the model to select only relevant features.

% Step 4: Hyperpriors for Population-Level Parameters

The precision parameters \( \boldsymbol{\alpha} = \{\alpha_j\} \) are given population-level Gamma hyperpriors to allow flexibility across users:
\[
P(\alpha_j) = \text{Gamma}(\alpha_j | a, b)
\]
where \( a \) and \( b \) are shape and rate parameters that encourage sparsity in the model.

% Step 5: Posterior Inference Using Bayesian Updating for Sequential Notes

Given the sequential nature of fragrance notes, we apply Bayesian updates to refine the preference prediction after each note:

1. Top Note:
   After observing the preference \( y_i^{(T)} \) for the top note, update the posterior for weights \( \mathbf{w} \):
   \[
   P(\mathbf{w} | \mathbf{y}^{(T)}, \mathbf{x}^{(T)}, \boldsymbol{\alpha}, \sigma^2) \propto P(\mathbf{y}^{(T)} | \mathbf{w}, \sigma^2) \cdot P(\mathbf{w} | \boldsymbol{\alpha})
   \]
   where \( \mathbf{y}^{(T)} \) and \( \mathbf{x}^{(T)} \) represent observed top note preferences and features, respectively.

2. Middle Note:
   Using the updated posterior from the top note, apply a second Bayesian update after observing the middle note preference \( y_i^{(M)} \):
   \[
   P(\mathbf{w} | \mathbf{y}^{(T)}, \mathbf{y}^{(M)}, \mathbf{x}^{(T)}, \mathbf{x}^{(M)}, \boldsymbol{\alpha}, \sigma^2) \propto P(\mathbf{y}^{(M)} | \mathbf{w}, \sigma^2) \cdot P(\mathbf{w} | \mathbf{y}^{(T)}, \mathbf{x}^{(T)}, \boldsymbol{\alpha}, \sigma^2)
   \]

3. Base Note:
   Finally, update after observing the base note preference \( y_i^{(B)} \), resulting in the final posterior:
   \[
   P(\mathbf{w} | \mathbf{y}^{(T)}, \mathbf{y}^{(M)}, \mathbf{y}^{(B)}, \mathbf{x}^{(T)}, \mathbf{x}^{(M)}, \mathbf{x}^{(B)}, \boldsymbol{\alpha}, \sigma^2) \propto P(\mathbf{y}^{(B)} | \mathbf{w}, \sigma^2) \cdot P(\mathbf{w} | \mathbf{y}^{(T)}, \mathbf{y}^{(M)}, \mathbf{x}^{(T)}, \mathbf{x}^{(M)}, \boldsymbol{\alpha}, \sigma^2)
   \]

% Step 6: Prediction of Final Fragrance Preference

After completing the updates for all notes, the final posterior is
\[
P(\mathbf{w} | \mathbf{y}^{(T)}, \mathbf{y}^{(M)}, \mathbf{y}^{(B)}, \mathbf{x}^{(T)}, \mathbf{x}^{(M)}, \mathbf{x}^{(B)}, \boldsymbol{\alpha}, \sigma^2) 
\] 
provides a personalized fragrance preference prediction for each user:
\[
\hat{y}_i = \mathbf{w}^T \boldsymbol{\phi}(\mathbf{x}_i)
\]
where \( \hat{y}_i \) is the predicted preference score based on the sequential experiences with the top, middle, and base notes.

This HBM using RVM mathematics provides a probabilistic, personalized fragrance preference model that combines user-specific data with population trends while incorporating sequential Bayesian updates for each note.

% Data Required for Estimating a Hierarchical Bayesian Relevance Vector Machine for Personalized Fragrance Preferences Based on Jungian Perspective

From a Jungian perspective, the data needed to estimate the relevance vector machine (RVM) in a hierarchical Bayesian algorithm for personalized fragrance preferences should capture both psychological traits and sensory responses. Jungian concepts like personality archetypes and cognitive functions can inform user preferences as they relate to fragrance characteristics. Here’s a breakdown of the essential data types:

% 1. Personality Archetypes and Cognitive Functions
\section{Personality Archetypes and Cognitive Functions}
\begin{itemize}
    \item \textbf{Jungian Archetypes}: Collect data on users’ dominant archetypes (e.g., Hero, Caregiver, Explorer), as these archetypes influence the types of scents users are likely to prefer. For example, an Explorer might prefer adventurous, earthy scents, while a Caregiver might lean toward soft, comforting fragrances.
    \item \textbf{Cognitive Functions}: Gather scores on each of the eight Jungian cognitive functions (e.g., Extraverted Thinking, Introverted Feeling). For example, those with a dominant \textit{Introverted Feeling} function might prefer subtle, intimate fragrances, while \textit{Extraverted Intuition} types may seek novel or complex scents.
\end{itemize}
These variables will be used as predictors in the RVM and can be extracted through self-assessment questionnaires, such as the Myers-Briggs Type Indicator (MBTI), which is inspired by Jung’s theories.

% 2. Sensory and Emotional Associations with Scents
\section{Sensory and Emotional Associations with Scents}
\begin{itemize}
    \item \textbf{Emotional Response Data}: Gather information on how users feel in response to specific scents. Emotions can be categorized as uplifting, calming, sensual, grounding, etc. Jung emphasized the psychological effects of sensory experiences, so emotional responses are central to understanding fragrance preferences.
    \item \textbf{Symbolic Associations}: Collect associations between scents and symbolic imagery (e.g., forest for wood scents, ocean for fresh notes). This connects to Jungian symbolism, as scents often evoke archetypal symbols that resonate differently with each personality type.
\end{itemize}

% 3. Top, Middle, and Base Note Preferences
\section{Top, Middle, and Base Note Preferences}
\begin{itemize}
    \item \textbf{Individual Preferences by Note Layer}: Track preferences for specific fragrance notes within each layer—top, middle, and base. For example, which top notes (citrus, floral), middle notes (spice, wood), or base notes (amber, musk) are preferred? Users’ Jungian personalities can influence these preferences; for instance, people with \textit{Sensation} as a primary cognitive function may favor earthy, grounded notes, while \textit{Intuitive} types may lean toward more abstract or ethereal scents.
    \item \textbf{Intensity and Complexity Preferences}: Record user preferences regarding scent intensity (light, moderate, strong) and complexity (single note, blended, layered). Jungian types oriented toward Extraverted Sensing might prefer intense fragrances, while Introverted Intuition types might enjoy more subtle blends.
\end{itemize}

% 4. Contextual Preferences and Environmental Factors
\section{Contextual Preferences and Environmental Factors}
\begin{itemize}
    \item \textbf{Situational Preferences}: Track preferences for different contexts or moods (e.g., work, social, relaxation). Jung’s theories on persona and shadow imply that fragrance preferences can shift depending on one’s social role or context, making it valuable to capture situational preferences.
    \item \textbf{Environmental and Cultural Influences}: Collect demographic and cultural data, as scent preferences often align with cultural values and environmental factors. Jung noted the role of collective consciousness, so understanding cultural or regional scent preferences can enhance the model’s relevance to diverse user groups.
\end{itemize}

% 5. Demographic Data
\section{Demographic Data}
Basic demographic information (age, gender, etc.) can help fine-tune the relevance of archetypes and scent preferences across broader population trends, aligning the RVM model with Jungian patterns that may vary by demographics.

% Structuring the Data for the RVM in an HBM
\section{Structuring the Data for the RVM in an HBM}
For the RVM’s hierarchical Bayesian structure, the data can be organized as follows:
\begin{itemize}
    \item \textbf{User-Level Inputs} (\( \mathbf{x}_i \)): Personality archetypes, cognitive functions, emotional responses, note preferences, and situational preferences.
    \item \textbf{Population-Level Inputs} (\( \mu, \sigma \)): Hyperparameters representing population trends across archetypes, collective preferences for certain scent types, and intensity/complexity patterns.
\end{itemize}

This data will allow the RVM-based hierarchical Bayesian model to learn the relevance of each feature, tailoring fragrance recommendations dynamically according to the user’s Jungian personality profile and sequential note experiences.

% Jungian Personality Archetypes

\section{Jungian Personality Archetypes}

Jungian personality archetypes are recurring, universal characters or motifs that represent distinct patterns in human behavior and personality. These archetypes can influence a user’s sensory preferences, including fragrance choices, by providing insights into their core motivations and emotional resonances. Below are some of the primary archetypes identified by Carl Jung that can be relevant in shaping fragrance preferences:

\begin{itemize}
    \item \textbf{The Hero}: Characterized by courage, ambition, and a desire for mastery. Heroes may prefer bold, invigorating scents that convey strength and resilience, such as woody, spicy, or intense citrus notes.

    \item \textbf{The Caregiver}: Associated with compassion, nurturing, and a sense of duty. Caregivers often gravitate toward soft, comforting fragrances, like warm vanilla, gentle florals, and mild, sweet aromas that evoke a sense of safety and warmth.

    \item \textbf{The Explorer}: Driven by a desire for freedom, discovery, and new experiences. Explorers may favor adventurous, earthy, or complex scents, including herbal, smoky, or natural tones that convey independence and connection to nature.

    \item \textbf{The Lover}: Focused on intimacy, passion, and sensuality. Lovers typically enjoy romantic, alluring fragrances such as deep florals (e.g., rose, jasmine), musks, and oriental blends that enhance their charismatic and intimate nature.

    \item \textbf{The Sage}: Pursues knowledge, truth, and wisdom. Sages might appreciate sophisticated, understated fragrances that reflect their intellectual nature, often preferring subtle, balanced notes such as green tea, cedarwood, and sandalwood.

    \item \textbf{The Jester}: Known for playfulness, joy, and spontaneity. Jesters may be drawn to fun, light, and refreshing scents, such as fruity or citrus-based notes, that match their lively and optimistic outlook.

    \item \textbf{The Ruler}: Desires control, order, and influence. Rulers often prefer classic, powerful fragrances like leather, tobacco, or amber that embody sophistication and authority.

    \item \textbf{The Innocent}: Values simplicity, purity, and optimism. Innocents tend to favor clean, fresh, and subtle scents, such as light florals, citrus, and soft green notes, which convey clarity and simplicity.

    \item \textbf{The Rebel (Outlaw)}: Seeks to challenge norms and values freedom and individuality. Rebels may be attracted to unconventional, intense scents, such as smoky, leathery, or metallic notes that reflect their defiance and edgy persona.

    \item \textbf{The Magician}: Focused on transformation, vision, and innovation. Magicians might appreciate mystical, layered scents like incense, oud, or complex floral arrangements that capture their fascination with the mysterious and the extraordinary.
\end{itemize}

Each of these archetypes represents a different facet of human personality, which can impact an individual’s scent preferences. By incorporating archetypes into the fragrance preference model, the hierarchical Bayesian RVM can use these psychological characteristics to make more nuanced and meaningful scent recommendations.

% Data Needed to Estimate Personality Based on Jungian Archetypes for a Hierarchical Relevance Vector Machine in Personalized Fragrance Preferences

To estimate each person's personality in terms of Jungian archetypes for a hierarchical relevance vector machine (RVM) applied to personalized fragrance preferences, we need data that quantifies how strongly each archetype is expressed within an individual. This data enables us to assign feature weights to different personality traits, which are used in the hierarchical Bayesian RVM model for sequential preference updating with the top, middle, and base notes.

% Step 1: Required Data for Archetype-Based Personality Estimation

For each individual \( i \), we aim to estimate a vector of archetype intensities, \( \mathbf{a}_i = [a_{i, \text{Hero}}, a_{i, \text{Caregiver}}, \ldots, a_{i, \text{Magician}}] \), where each element represents the degree to which a specific Jungian archetype is expressed. The following data types are required to estimate \( \mathbf{a}_i \):

1. Self-Reported Questionnaire Data:
   \begin{itemize}
       \item A structured questionnaire assessing personality traits aligned with each archetype. Examples of scales include:
           \[
           Q_{i, \text{Hero}}, Q_{i, \text{Caregiver}}, \ldots, Q_{i, \text{Magician}}
           \]
       where \( Q_{i, j} \) is a score for individual \( i \) on questions corresponding to archetype \( j \). These scores can be transformed into archetype intensities \( a_{i,j} \) after normalization and scaling.
   \end{itemize}

2. Behavioral Data:
   \begin{itemize}
       \item Observations of actions or preferences that align with archetypal behavior, such as choices in products, hobbies, or activities. Behavioral data is mapped to each archetype as binary indicators or scaled scores, which can be incorporated into the model as:
           \[
           B_{i, j} \sim \text{Bernoulli}(\pi_{i,j}) \quad \text{or} \quad B_{i, j} \sim \mathcal{N}(\mu_{i,j}, \sigma_{i,j}^2)
           \]
       where \( B_{i, j} \) is an indicator for behavior \( i \) aligning with archetype \( j \).
   \end{itemize}

3. Demographic and Contextual Data:
   \begin{itemize}
       \item Demographic factors (e.g., age, gender, culture) that influence archetypal tendencies in scent preferences. These variables provide context and weight to certain archetypes, enhancing accuracy for hierarchical estimation. Let:
           \[
           D_{i} = [\text{age}_i, \text{gender}_i, \text{culture}_i, \ldots]
           \]
       where each component can contribute a prior influence on the archetypes.
   \end{itemize}

% Step 2: Structuring Data into the Hierarchical Model

Each individual’s archetype profile is a vector \( \mathbf{a}_i \), derived from the above data sources, and represents the prior distribution for their scent preferences. To model fragrance preferences hierarchically, we represent individual preferences as follows:

1. Individual-Level Model for Archetypes and Preferences:
   \[
   y_i \sim \mathcal{N}(\mathbf{w}^T \boldsymbol{\phi}(\mathbf{a}_i), \sigma^2)
   \]
   where:
   \begin{itemize}
       \item \( y_i \): Overall fragrance preference score for individual \( i \).
       \item \( \boldsymbol{\phi}(\mathbf{a}_i) \): Basis function mapping archetype intensities \( \mathbf{a}_i \) to a high-dimensional space capturing non-linear relationships in fragrance preferences.
       \item \( \mathbf{w} \): Weight vector, representing the relevance of each archetype for fragrance preference prediction.
   \end{itemize}

2. Population-Level Priors for Archetypes:
   We define population-level priors for each archetype, which are shared across individuals but adjusted by demographic variables. These priors allow for a population-level influence that guides individual archetype distributions:
   \[
   \mathbf{a}_i \sim \mathcal{N}(\boldsymbol{\mu}_{\mathbf{a}} + \mathbf{\beta}^T \mathbf{D}_i, \boldsymbol{\Sigma}_{\mathbf{a}})
   \]
   where:
   \begin{itemize}
       \item \( \boldsymbol{\mu}_{\mathbf{a}} \): Population mean vector for archetype intensities.
       \item \( \mathbf{D}_i \): Demographic vector for individual \( i \).
       \item \( \mathbf{\beta} \): Coefficient matrix aligning demographic factors with archetype intensities.
       \item \( \boldsymbol{\Sigma}_{\mathbf{a}} \): Covariance matrix representing variability across the population.
   \end{itemize}

% Step 3: Hierarchical Bayesian Updating with Fragrance Notes

We apply Bayesian updating to refine preferences after each fragrance note layer:

1. Top Note Update:
   After observing the preference \( y_i^{(T)} \) for the top note, we update the posterior of weights \( \mathbf{w} \) based on individual and population-level archetype data:
   \[
   P(\mathbf{w} | y_i^{(T)}, \mathbf{a}_i, \mathbf{D}_i, \boldsymbol{\mu}_{\mathbf{a}}, \boldsymbol{\Sigma}_{\mathbf{a}}, \sigma^2) \propto P(y_i^{(T)} | \mathbf{w}, \sigma^2) \cdot P(\mathbf{w} | \mathbf{a}_i)
   \]

2. Middle Note Update:
   With updated preferences from the top note, a second Bayesian update follows after the middle note preference \( y_i^{(M)} \):
   \[
   P(\mathbf{w} | y_i^{(T)}, y_i^{(M)}, \mathbf{a}_i, \mathbf{D}_i, \boldsymbol{\mu}_{\mathbf{a}}, \boldsymbol{\Sigma}_{\mathbf{a}}, \sigma^2) \propto P(y_i^{(M)} | \mathbf{w}, \sigma^2) \cdot P(\mathbf{w} | y_i^{(T)}, \mathbf{a}_i)
   \]

3. Base Note Update:
   After observing the base note preference \( y_i^{(B)} \), the final posterior reflects all note experiences:
   \[
   P(\mathbf{w} | y_i^{(T)}, y_i^{(M)}, y_i^{(B)}, \mathbf{a}_i, \mathbf{D}_i, \boldsymbol{\mu}_{\mathbf{a}}, \boldsymbol{\Sigma}_{\mathbf{a}}, \sigma^2) \propto P(y_i^{(B)} | \mathbf{w}, \sigma^2) \cdot P(\mathbf{w} | y_i^{(T)}, y_i^{(M)}, \mathbf{a}_i)
   \]

% Final Posterior Preference

After sequential updates, the final posterior, \( P(\mathbf{w} | y_i^{(T)}, y_i^{(M)}, y_i^{(B)}, \mathbf{a}_i, \mathbf{D}_i, \boldsymbol{\mu}_{\mathbf{a}}, \boldsymbol{\Sigma}_{\mathbf{a}}, \sigma^2) \), provides a tailored fragrance profile for individual \( i \), integrating both individual archetype data and population-level trends.

\bibliographystyle{plainnat}
\bibliography{references}

\end{document}